\documentclass[10pt,aps,prl,showpacs,twocolumn]{revtex4-2}
\usepackage{amsmath, amssymb}
\usepackage{amsthm}
\usepackage{graphicx}
\usepackage{subcaption}
\usepackage{mathtools}

\usepackage[T1]{fontenc}

\newtheorem{thrm}{Theorem}

\renewcommand{\vec}[1]{\mathbf{#1}}

\begin{document}
	
\title{Optimal entanglement swapping in quantum repeaters}
\author{Evgeny Shchukin}
\email{evgeny.shchukin@gmail.com}
\author{Peter van Loock}
\email{loock@uni-mainz.de}
\affiliation{Johannes-Gutenberg University of Mainz, Institute of Physics, Staudingerweg 7, 55128 Mainz, Germany}

\begin{abstract}
We formulate the problem of finding the optimal entanglement swapping scheme in a quantum repeater chain as a Markov
decision process and present its solution for different repeater's sizes. Based on this, we are able to demonstrate that
the commonly used ``doubling'' scheme for performing probabilistic entanglement swapping of probabilistically
distributed entangled qubit pairs in quantum repeaters does not always produce the best possible raw rate. Focussing on
this figure of merit, without considering additional probabilistic elements for error suppression such as entanglement
distillation on higher ``nesting levels'', our approach reveals that a power-of-two number of segments has no privileged
position in quantum repeater theory; the best scheme can be constructed for any number of segments. Moreover, classical
communication can be included into our scheme, and we show how this influences the raw waiting time for different number
of segments, confirming again the optimality of ``non-doubling'' in some relevant parameter regimes. Thus, our approach
provides the minimal possible waiting time of quantum repeaters in a fairly general physical setting.
\end{abstract}


\maketitle

\textit{Introduction}. A long-standing problem in the theory of quantum repeaters is to determine the optimal
entanglement distribution time as a function of a repeater's characteristics like the distribution success probability
$p$ for a single segment and the entanglement swapping probability $a$ between two segments \footnote{In a typical
quantum repeater system, the parameter $p$ is primarily given by the probability that a photonic qubit is successfully
transmitted via a fiber channel of length $L_0$ connecting two stations, $e^{-L_0/22\mathrm{km}}$. It also includes
local state preparation/detection, fiber coupling, frequency conversion, and memory write-in efficiencies. The parameter
$a$ can be related to the memory read-out or an optical Bell measurement efficiency.}. In particular, it is commonly
assumed that the number of segments $n$ is a power of two, $n = 2^d$, and the only scheme considered is ``doubling'',
where the segments are divided into two equal halves, which are then treated as independent smaller repeaters. When both
halves have finally distributed an entangled state the last swapping is attempted. Such a doubling scheme can be useful
in a ``nested quantum repeater'' allowing for a systematic inclusion of entanglement distillation
\cite{PhysRevLett.81.5932} or, exploiting the repeater's ``self-similarity'', for a recursive and approximate
calculation of repeater rates in certain regimes \cite{RevModPhys.83.33}. However, it is unknown whether ``doubling''
gives the optimal rates, i.e. the shortest repeater waiting times -- in general, for general $(n = 2^d, p, a)$ without
entanglement distillation, or at least in a certain regime such as that of $p$ and $a$ both being small. Furthermore,
one may ask: is there an optimal scheme if $n \neq 2^d$?

Rate calculations for quantum networks so far have focussed either on the ultimate, information-theoretical limits
independent of experimental constraints such as non-deterministic gate operations \cite{CommPhys.2.51} or on more
realistic systems under simplifying assumptions, such as specific parameter regimes allowing for an approximate
treatment \cite{RevModPhys.83.33} or to determine bounds \cite{coopmans2021improved} and certain shapes
\cite{10.1145/3374888.3374899} and sizes \cite{Luong2016, PhysRevA.99.052330} of the network reducing its complexity. In
this work, we bridge these two approaches for the case of a sufficiently small quantum repeater chain up to about ten
segments and present its exact, optimal solutions, generalizing and optimizing our previous, exact results on the
statistics of repeater waiting times in various settings \cite{PhysRevA.100.032322} \footnote{See also
\cite{PhysRevA.99.042313} and \cite{10.1117.12.811880}. In our previous work \cite{PhysRevA.100.032322}, we have shown
how the exact repeater rates can be obtained for arbitrary $(n, p, a)$ and arbitrary fixed (e.g. ``doubling'') or
dynamical (e.g. ``swap as soon as possible'') protocols. However, even including methods compressing the Markov state
space (``lumpability''), the resulting set of linear equations to be solved grows exponentially with the size of the
repeater $n$, and so explicit expressions can be given in a compact form only for smaller repeaters such as $n=4$; for
larger repeaters, the linear system must be solved directly. Note that in terms of their possible Markov states, a
quantum repeater chain is remarkably complex. The rates of other classes of quantum networks, for example, a star-shaped
``entanglement switch'' with a single center node \cite{10.1145/3374888.3374899} may be computed more easily and
efficiently (e.g. by counting the number of successfully distributed segments where actions like Bell measurements do
not depend on the availability of specific segments \cite{10.1145/3374888.3374899}).}.

It turns out that, depending on $p$ and $a$, the ``doubling'' scheme does not always deliver the highest raw rate, and
for some values of parameters other schemes perform better. The corresponding rate enhancement seems to increase for
larger repeaters with growing $n = 2^d$. Moreover, the assumption that $n$ is a power of two is superfluous, the optimal
scheme is defined for all combinations of $n$, $p$, and $a$ \footnote{If memory and gate errors are included and state
fidelities or bit error rates in quantum key distribution are considered, it depends on the corresponding hardware
parameter regime whether ``doubling'' with entanglement distillation or swapping as soon as possible performs best,
which can be found through numerical simulations (see \cite{CommPhys.4.164} and also references therein). The present
work instead shows via rigorous and exact analytical optimizations of statistical quantities in a physical setting that
even with sufficiently good hardware, but assuming non-deterministic entanglement swapping (which actually may come
along with an imperfect lossy memory system such as a collective bosonic spin mode of an atomic ensemble
\cite{Nature.414.413}), a different performance ordering of the various protocols can occur depending on the parameter
triple $(n, p, a)$. For a nice review on the different approaches to treat quantum repeaters and networks
quantitatively, either analytically under ideal or more realistic conditions or, even more realistically but by omitting
the possible general insights that primarily an analytical treatment may provide, via numerical simulations, see
\cite{doi:10.1116/5.0024062}.}. In this work, we show how this scheme can be found. We will also include the physically
relevant case where the memory qubits have to wait for classical signals to obtain information regarding the
distributions in other segments \footnote{The classical communication times are an important limiting factor in a
realistic quantum network and the corresponding schemes are hard to optimize. Unlike previous treatments of nested
``doubling'' schemes based on approximations for small $p$ and $a$ \cite{PhysRevA.87.062335} we exactly derive the
optimal schemes with classical communication independent of the $n = 2^d$ assumption for any $(n, p, a)$ up to about
$n=10$.}. We assume that all segments have identical properties \footnote{Similarly, all intermediate stations can
perform entanglement swapping with the same probability. Considering equal segment lengths $L_0$ is then a reasonable
assumption for a linear quantum repeater chain. The ultimate repeater-assisted capacity \cite{CommPhys.2.51} is
determined by the capacity of the longest segment and hence is also maximized for a repeater with equal segment
lengths.} and put no restriction on how long a state can be kept in memory \footnote{The schemes we consider for our
rate optimizations are error-free. In particular, memory qubits can be held as long as needed. However, we do not assume
an unlimited availability of quantum memories. If a segment is filled with a single entangled pair there are no further
distribution attempts in that segment which is a common assumption \cite{RevModPhys.83.33, Luong2016,
PhysRevA.99.052330, PhysRevA.100.032322}. While it can be useful to continue distributing and load additional memory
qubits or just replace an already held one by a fresh, newly distributed pair (in case of a memory state fidelity
decaying with time), provided the extra resources are available, the corresponding (secret key) rate analysis gets again
more involved requiring simplifying assumptions on the network topology to allow for more general ``memory buffers''
\cite{10.1145/3374888.3374899} or a numerical treatment \cite{CommPhys.4.164}. Our schemes have unit buffer size and it
would not affect the raw rates if we continued distributing pairs in a loaded segment to replace the existing pair. An
infinite buffer size makes most efficient use of the quantum communication channel. In this kind of unphysical case,
assuming unlimited resources, the optimal rates can be calculated even for a repeater chain and it turns out that
``doubling'' is always optimal when $n = 2^d$ \cite{8967073}. Besides memory buffers, a ``memory cutoff'' can be
introduced where a memory state is discarded after waiting for too long \cite{PhysRevLett.98.060502}, reducing the raw
communication rates but preventing state fidelities from dropping below a predetermined value. In the exact Markov chain
approach, a finite memory cutoff can be included at the expense of computational efficiency due to a further increase of
the linear equation system \cite{PhysRevA.100.032322}. In principle, it could also be included into the optimizations of
the present work. Finally, simple multiplexing with several repeater chains \cite{PhysRevLett.98.060502} could also be
incorporated into our schemes by replacing the corresponding $p$ value in each segment.}.

\textit{Markov chains}. We use the formalism of Markov decision processes, which provides a method to choose an optimal
action in a discrete-time stochastic system. Before we formulate our general method, we solve a simpler problem.
Consider a finite Markov chain with a single absorbing state. The set of states we denote as $\mathcal{S}$ and the
transition probability matrix as $P = (p_{ss'})$, where $p_{ss'}$ is the transition probability from $s$ to $s'$. With
every state $s \in \mathcal{S}$ we associate a cost $r_s \geqslant 0$ of making a transition from this state. Assuming
that the cost of the absorbing state is zero, we define the total cost of absorption $T_s$ from any state $s \in
\mathcal{S}$ as a sum of all costs $r_{s'}$ from $s$ to the absorbing state. Clearly, $T_s$ is a random variable whose
distribution depends on the transition probabilities of the chain. What is the average value of this variable? Denoting
$v_s = \mathbf{E}[T_s]$, it can be shown that these quantities satisfy the system of linear equations
\begin{equation}\label{eq:v2}
    v_s = \sum_{s' \in \mathcal{S}}p_{ss'}v_{s'} + r_s.
\end{equation}
Following the convention that the absorbing state is the last one let $Q$ be the stripped $P$, i.e. the matrix obtained
from $P$ by removing its last row and last column. The system \eqref{eq:v2} then reads as $\vec{v} = Q\vec{v} +
\vec{r}$, and its solution is given by $\vec{v} = (I - Q)^{-1}\vec{r}$, where $I$ is the identity matrix of order $n-1$,
$n = |\mathcal{S}|$, and $\vec{r}$ is the $(n-1)$-vector of transition costs (except the last component, which we
assumed to be zero). It is known that $I-Q$ is invertible and thus the system \eqref{eq:v2} has a unique solution given
by this expression, see \cite{PhysRevA.100.032322}. This formula gives an exact analytical expression for the average
absorption cost. For large $n$ such an expression is impractical to deal with, so $\vec{r}$ must be computed
numerically. In this case solving the system of linear equations $(I-Q)\vec{v} = \vec{r}$ is numerically more robust
than inverting the matrix $I - Q$ and multiplying the inverse by $\vec{r}$.

\textit{Markov decision problems}. Now consider a more complicated case. What if the transition probabilities and the
associated transition cost in each state depend on a parameter, so-called action? These actions can be freely chosen at
will and any such a choice is referred to as a policy in the reinforcement learning parlance. A (finite) Markov decision
process (MDP) consists of a finite set $\mathcal{S}$ of states and a finite set of actions $\mathcal{A}_s$ for each
state $s \in \mathcal{S}$. For every policy we have an instance of the problem considered above, with the transition
probabilities $p^\alpha_{ss'}$ and transition costs $r^\alpha_s$ now depending on the actions $\alpha \in \mathcal{A}_s$
chosen for all $s \in \mathcal{S}$. One can say that an MDP embeds many Markov chain absorption problems into one
framework. What is the optimal (minimal or maximal) average absorption cost of these problems and how can we determine
the corresponding optimization problem? We are interested in minimizing the cost, but all the results below can be
easily modified for the opposite direction.

A straightforward approach is to compute the average cost of all embedded problems \eqref{eq:v2} and take the best
value. Totally, there are $N = \prod_{s \in \mathcal{S}} |\mathcal{A}_s|$ embedded problems, and this number becomes
ridiculously large even for problems of moderate size, so this method is feasible only for very small systems (see SM).
We show that there is a more practical approach based on solving a proper linear optimization problem. This statement is
based on the following
\begin{thrm}
Any solution of the linear optimization problem which maximizes the sum $\sum_{s \in \mathcal{S}} v_s$ under the
constraints
\begin{equation}\label{eq:of}
    v_s \leqslant \sum_{s' \in \mathcal{S}} p^\alpha_{ss'} v_{s'} + r^\alpha_s, \quad \alpha \in \mathcal{A}_s
\end{equation}
is a solution of the following system of nonlinear equations:
\begin{equation}\label{eq:min}
    v_s = \min_{\alpha \in \mathcal{A}_s} \biggl[\sum_{s' \in \mathcal{S}} p^\alpha_{ss'} v_{s'} + r^\alpha_s \biggr].
\end{equation}
As the objective function one can use any linear combination $\sum_{s \in \mathcal{S}} c_s v_s$ with positive
coefficients $c_s$. The problem \eqref{eq:of} has at least one solution.
\end{thrm}
The proof of this theorem is given in the Supplemental Material. If for any concrete choice of $\alpha \in
\mathcal{A}_s$ for all $s \in \mathcal{S}$, i.e. for any policy $\pi$, we leave just one equation in Eq.\eqref{eq:min},
we get a system of linear equations of the form given by Eq.~\eqref{eq:v2}. The solution $\vec{v}^\pi$ of this system is
the vector of average costs of the absorption problem corresponding to the policy $\pi$. It is in this sense that an MDP
embeds many absorption problems --- every choice of actions produces a problem and all these problems are contained in
one framework described by Eq.~\eqref{eq:min}. Note that any solution of Eq.~\eqref{eq:min} (which has at least one
solution according to the previous theorem) corresponds to a policy --- for any $s \in \mathcal{S}$ take an action
$\alpha \in \mathcal{A}_s$ that minimizes the right-hand side of Eq.~\eqref{eq:min}. For some $s$ there can be more than
one minimizing action, so the policy corresponding to a solution may not be unique. We now show that any solution of
Eq.~\eqref{eq:min} is at least as optimal as the solution for any policy.

\begin{thrm}
Let $\vec{v}^*$ be a solution of Eq.~\eqref{eq:min}. Then for any policy $\pi$ we have $\vec{v}^* \leqslant
\vec{v}^\pi$, where this inequality is meant componentwise.
\end{thrm}
The proof is also given in the Supplemental Material. As it was noted before, any optimal solution corresponds to some
policy, so from this theorem we derive the following property of optimal solutions: $v^*_s = \min_\pi v^\pi_s$, where
the minimum is taken over all possible policies. We conclude that the system of nonlinear equations Eq.~\eqref{eq:min}
has a unique solution, which can be obtained by solving the linear optimization problem given by Eq.~\eqref{eq:of}.
Having found the optimal solution $\vec{v}^*$ we can obtain an optimal policy corresponding to this solution by taking
an action $\alpha \in \mathcal{A}_s$ that minimizes the right-hand side in Eq.~\eqref{eq:min} for all $s \in
\mathcal{S}$. Such a scheme may not be unique.

\textit{Application to quantum repeaters}. We now apply the presented theory to the problem of finding the minimal
waiting time in quantum repeaters. In a state where there are segments not ready yet (which are trying to distribute an
entangled state) and those that are ready (which have already distributed entanglement), there is always a choice ---
either wait for non-ready segments or try to swap a pair of neighboring ready segments. Clearly, different actions have
different probabilistic evolutions, so the entanglement distribution process in a quantum repeater fits into an MDP
model.

First, we need to list all possible states of a quantum repeater. We use a simple model where an attempt to distribute
entanglement takes one unit of time and an attempt to swap segments takes no time at all. Under these assumptions a
state of a repeater can be characterized by a string of nonnegative numbers, where 0 marks a segment trying to
distribute entanglement, and $i>0$ marks a group of $i$ successfully distributed and swapped segments. For the simplest
case of a 2-segment repeater the states are $00$ (the initial state), $01$, $10$, $11$ and $2$ (the final, absorbing
state). We are interested in an optimal strategy, and such a strategy must have identical actions on the states which
are mirror images of each other, like the states $01$ and $10$ above. It means that we can apply the so-called
lumpability trick --- we lump the mirror images into one new state and recompute the transition probabilities. This
allows us to compress the size of the problem by reducing the number of states and actions, which will be very helpful
for larger repeaters. In the case above from the two states $01$ and $10$ we form a new state $\{01, 10\}$. So, in this
simple case we have four states $s_1 = 00$, $s_2 = \{01, 10\}$, $s_3 = 11$ and $s_4 = 2$. In each of these states only
one action is possible, so our MDP reduces to the Markov chain problem of the form \eqref{eq:v2}:
\begin{equation}\label{eq:r2}
\begin{split}
    v_1 &= q^2 v_1 + 2pq v_2 + p^2 v_3 + 1 \\
    v_2 &= q v_2 + p v_3 + 1 \\
    v_3 &= (1-a) v_1 + a v_4,
\end{split}
\end{equation}
where $v_4 = 0$ (and we set $q=1-p$). Note that the transition probability $p_{12} = \mathbf{P}(s_1 \to s_2) = 2pq$ has
a factor 2, since $s_2 = \{01, 10\}$ and $s_1 = 00$ can go to $s_2$ in two ways --- when either of the segments
distributes entanglement. The probability of each path is $pq$, so the total transition probability is $2pq$. The
constant terms on the right-hand side of the system \eqref{eq:r2} express our assumption that a distribution attempt
costs one unit of time and a swapping attempt costs zero. Solving this system of linear equations, we obtain $v_1 =
(3-2p)/(ap(2-p))$, which is a well-known expression for the waiting time of a 2-segment repeater. Note that for the
total repeater waiting time we are generally interested in component $v_1$ from the optimal solution vector.

\textit{Example}. Now consider a more interesting case, a 3-segment repeater. In this case there are nine states: $s_1 =
000$, $s_2 = 001$, $s_3 = 010$, $s_4 = 011$, $s_5 = 101$, $s_6 = 111$, $s_7 = 02$, $s_8 = 12$ and $s_9 = 3$, where any
non-symmetric sequence like $001$ denotes the corresponding class $\{001, 100\}$ not to overload the notation. In the
state $s_4 = 011$ (which denotes $\{011, 110\}$) two actions are possible --- waiting while the last segment distributes
entanglement, which costs one time unit per attempt, or trying to swap the other two segments, which costs nothing (in
the state $111$ the two possible swappings represent one action in the compressed system). The MDP equations in this
case read as
\begin{displaymath}
\begin{split}
    v_1 &= q^2(qv_1 + 2p  v_2 + p v_3) + p^2(2qv_4 + qv_5 + pv_6) + 1 \\
    v_2 &= q^2 v_2 + pqv_4+pqv_5 + p^2 v_6 + 1 \\
    v_3 &= q^2 v_3 + 2pq v_4 + p^2 v_6 + 1 \\
    v_4 &\leqslant q v_4 + p v_6 + 1 \\
    v_4 &\leqslant (1 - a)v_1 + a v_7 \\
    v_5 &= qv_5 + pv_6 + 1 \\
    v_6 &= (1-a)v_2 + a v_8 \\
    v_7 &= q v_7 + p v_8 + 1 \\
    v_8 &= (1 - a)v_1,
\end{split}
\end{displaymath}
where we took into account that $v_9 = 0$. If we remove the first inequality for $v_4$, we get the scheme where we
always swap in the states $011$ and $110$, removing the second inequality we get the scheme where we always try to
distribute entanglement in these states \footnote{For the present $n=3$ case, there is only one state, $s_4$, that
allows for more than a single action, namely two, and so we have two inequalities for $v_4$. For the optimal solution,
one becomes an equation (the one corresponding to swapping), the other (the one corresponding to distributing) can be
removed, and the remaining system has 8 linear equations for 8 unknowns. This system of linear equations corresponds to
a Markov chain, like in Eq.~\eqref{eq:v2}, and from any state we can go to some other states with some probabilities.
However, prior to optimization, we have several actions, and the probabilities depend on the action chosen. In this
sense generally the MDP embeds many Markov chains and only when we choose one action for each state, we get back a
standard Markov chain. The relevant question here is what actions to choose to get the optimal behavior.}. Optimizing
the sum $\sum^8_{i=1}v_i$ under the constraints given above, for each $p$ and $a$ we obtain the best waiting time
$v^*_1$ and the optimal scheme (which may depend on $p$ and $a$). It turns out that for all $p$ and $a$ the former
scheme (always swapping when ready) is better and the analytical expression for the waiting time is the same as we have
already given in \cite{PhysRevA.100.032322}, where it is denoted as $\overline{K}^{(\mathrm{dyn})}_3$.

\begin{figure}
    \includegraphics[scale=0.9]{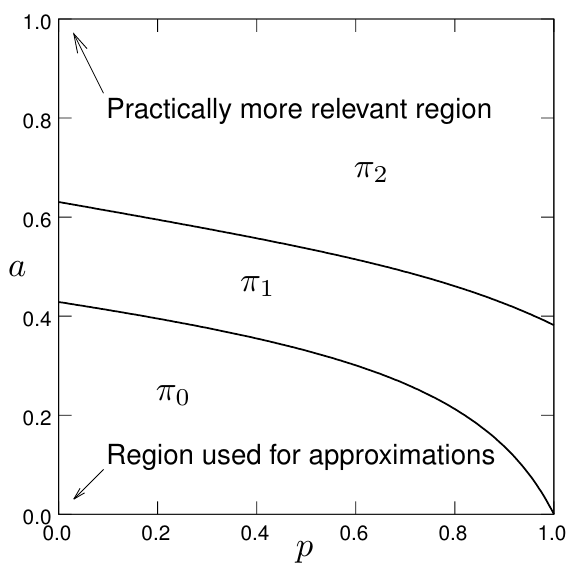}
    \caption{Regions of different optimal schemes, $n=4$.}\label{fig:BS4}
\end{figure}

\textit{Non-doubling optimal schemes}. The next case of a 4-segment repeater is even more interesting. The corresponding
MDP has 20 variables (excluding the variable for the absorbing state, whose value is zero) and 29 constraints, so we do
not present it explicitly. For every state there is at most one ``wait for distribution'' action and zero or more
``swapping'' actions. One of the possible schemes is ``doubling'', where the repeater is divided into two halves which
are treated as independent 2-segment repeaters. Only when both halves are ready can we try to perform the last swapping.
This scheme has been most commonly considered in the literature. Somewhat surprisingly, this scheme is not always the
best one. We have solved the MDP for all $0.01 \leqslant p, a \leqslant 1.0$ and for each pair of probabilities $p$ and
$a$ we determined the best scheme for these parameters. We found that in this case of $n=4$ there are three schemes that
are optimal in different regions of the probability square, see Fig.~\ref{fig:BS4}.

\begin{figure}
    \includegraphics{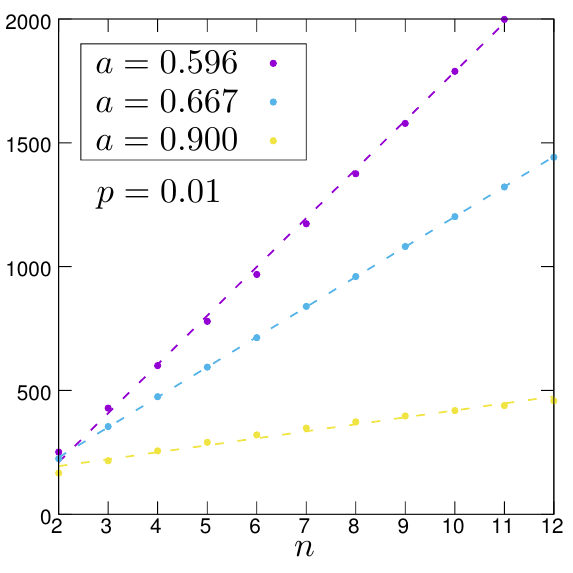}
\caption{Optimal raw waiting time as a function of $n$. For some values of $p$ and $a$ it is approximately linear.}
\label{fig:CC}
\end{figure}

\begin{figure}
    \includegraphics{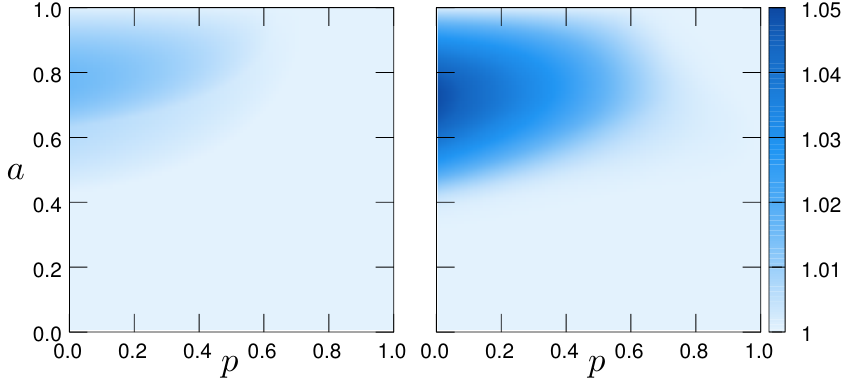}
\caption{Ratio of the ``doubling'' to the optimal waiting time for $n = 4$ (left) and $n = 8$ (right).}
\label{fig:OD4}
\end{figure}

In the lower-left corner of the square, which corresponds to small $p$ and $a$, the optimal scheme is ``doubling'',
denoted as $\pi_0$ (thus confirming that for such quantum repeaters ``doubling'' is indeed optimal). A practically more
relevant range of parameters may be at small $p$ and large $a$, which corresponds to the upper-left corner of the
square, and the optimal scheme there differs from ``doubling'', denoted as $\pi_2$. In between these two regions there
is a third optimal scheme, $\pi_1$. These schemes are described in the Supplemental Material. For some relevant fixed
$p$ and $a$ values, Fig.~\ref{fig:CC} illustrates that in an intermediate regime of $a$ (neither too small nor too close
to one), the optimized raw waiting time is a linear function of the repeater size $n$ \footnote{The value $p=0.01$
corresponds to a segment length of $L_0=100$km, only considering fiber channel loss. Thus, through segment numbers
$n=2...12$ we can cover total distances from 200km to 1200km. A swapping probability value of, for example, $a=2/3$ is
meaningful, because it corresponds to the efficiency of a linear-optics Bell measurement enhanced by means of 4
auxiliary photons \cite{PhysRevLett.113.140403}.}.

\textit{Classical communication}. We can extend our model to include classical communication (CC), assuming that it
takes one unit of time to restart a segment (and the swapping process itself takes no time). For example, for $n=4$ in
the state $0110$ we can try to swap the inner pair of segments. If the swapping fails, then in the previous model the
system transitions to the initial state $0000$, but in this model it goes to a new state $0(1)(1)0$, where the number in
brackets denotes the number of time units after which this segment returns to the initial state $0$. If we make a
swapping in the state $012$ and fail, this state goes to $0(1)(1)(2)$. With probability $q$ the next state is $000(1)$,
and with probability $p$ it is $100(1)$. In the former case the next state will be $***0$, where $***$ is any
combination of three zeros and ones, and in the latter case it will be $1**0$. So, the general rule is: $(i) \to (i-1)$
if $i>1$ and $(1) \to 0$. Let us illustrate possible transitions from the initial state $13$: $13 \to (1)(1)(2)(3) \to
00(1)(2) \to 110(1) \to (1)(1)0(1) \to 0000$. We first try to swap and fail, restarting all the segments. Then, two
segments are in the initial state and the other two are in progress still waiting for classical signals. Next, the ready
segments both distribute entanglement, simultaneously succeeding here, another segment goes to the initial state and the
last segment is still in progress. Then, we try to perform swapping and fail, restarting the first two segments (the
other two are in the same state since swapping itself takes no time). Finally, all segments are in the initial state,
since the third segment failed to distribute entanglement. Note that this is only one of the possible transition
sequences between repeater states in our model. These transitions illustrate that in multisegment repeaters several
``waves'' of restarting are possible --- an earlier restart signal still in progress when a newer one starts to
propagate. There are many more states and transitions by comparison with the previous model, but it still fits into the
MDP approach. The influence of the classical communication on the raw waiting time is illustrated in the Supplementary
Material for various repeater sizes \footnote{The typical repeater regime is at small $p$ where the rates are generally
low and so the relative impact of CC is small. However, note that the waiting times here are the inverse rates per total
channel use, independent of the actual duration of one time unit, i.e. the combined CC and local processing time per
segment (due to the blocking of successfully distributed segments, the total channel use is overestimated in our model).
As small $p$ values correspond to larger segment lengths and so larger CC times, the rates per second drop significantly
when including CC. Growing CC times also have a negative effect on the quantum memories which must store qubits for
longer.}. It is interesting to compare the optimal waiting time with the ``doubling'' waiting time. The ratio of the two
quantities is shown in Fig.~\ref{fig:OD4} for the model including CC. This figure shows that there is a small but
noticeable advantage of the optimal scheme. This advantage becomes more visible for larger repeaters, as
Fig.~\ref{fig:OD4}(right) demonstrates for an 8-segment repeater, in the practically highly relevant regime of small $p$
and large $a$ ($\sim$1.5\% for $n=4$ and $\sim$5\% for $n=8$).

\textit{Conclusion}. In conclusion, we presented a method to determine the most efficient entanglement swapping scheme
in a quantum repeater and demonstrated that the ``doubling'' scheme is not always the best. Moreover, our approach shows
that when additional elements such as entanglement distillation on higher levels are excluded the power-of-two number of
segments is not a distinguished case, since the best scheme can be constructed for any number of segments. We showed
that for small repeaters the best scheme has a tiny, but noticeable advantage over the ``doubling'' scheme, but this
advantage seems to increase with the repeater's size. Our most general model leading to this conclusion includes all
necessary classical communication times, while we were able to treat repeater sizes up to the order of ten segments. It
is currently intractable to treat 16 repeater segments or more for a direct comparison. 

Our algorithm has exponential complexity and thus is applicable to fairly ``small'' repeaters only, but even these
repeaters are still beyond current technological capabilities and so our approach here is fully applicable to current
experiments to have meaningful physical benchmarks. Moreover, a ten-segment repeater can cover a distance of around
1000km, which is already of practical interest. On the other hand, great progress has been made in algorithms for
solving linear optimization problems. A study of different versions of an optimization software, CPLEX, performed in
Ref.~\cite{Bixby2012ABH}, shows a speedup of a factor of 29,000 only due to algorithmic advantages. Combined with
hardware advances that happened during this time (two decades), we get an even more impressive performance boost factor.
Of course, no technological advance can turn an exponential algorithm to a subexponential one, but what seems
intractable now could become feasible in the near future.

\begin{acknowledgments}
We thank the BMBF in Germany for support via Q.Link.X and the BMBF/EU for support via QuantERA/ShoQC.
\end{acknowledgments}

\appendix

\section{Proof of Theorem 1}

We start with the proof of Theorem 1. Let $v^*_s$ be a solution of the optimization problem \eqref{eq:of}. We have to
prove that for any $s \in \mathcal{S}$ at least one of the inequalities \eqref{eq:of} for $\alpha \in \mathcal{A}_s$ is
equality. Let us assume that for some state $s_0$ all these inequalities are strict for the solution $v^*_s$. These
inequalities can be written as
\begin{equation}
    (1 - p^\alpha_{s_0 s_0})v^*_{s_0} < \sum_{s' \not= s_0} p^\alpha_{s_0 s'} v^*_{s'} + r^\alpha_{s_0},
\end{equation}
for all $\alpha \in \mathcal{A}_{s_0}$. Due to the relation $1 - p^\alpha_{s_0 s_0} \geqslant 0$ we can slightly
increase $v^*_{s_0}$ without violating these inequalities (and the inequalities for $s \not= s_0$ too, since $v^*_{s_0}$
appears there on the right-hand side with a nonnegative coefficient), so we can get a larger value of the objective
function $\sum_{s \in \mathcal{S}} c_s v_s$, which contradicts the assumption that $v^*_s$ is the optimal solution (it
is at this point that we use the condition $c_{s_0} > 0$). This proves that $v^*_s$ is a solution of Eq.~\eqref{eq:min}.

Now we prove that the optimization problem \eqref{eq:of} has a solution. We need to prove that the constraints are
feasible and that the objective function is bounded under the given constraints. The former statement is trivial, since
the point $\vec{v} = \vec{0}$ satisfies all the constraints, so we need only to prove the boundness. Because all sets of
actions $\mathcal{A}_s$ are non-empty, we can arbitrarily take an action for each state, which is equivalent to choosing
some scheme $\pi$. Taking only the conditions corresponding to the chosen actions in the system \eqref{eq:of}, we see
that any feasible point $\vec{v}$ satisfies the inequalities $\vec{v} \leqslant Q_\pi \vec{v} + \vec{r}_\pi$, where
$Q_\pi$ is obtained from $Q$ by leaving the rows corresponding to the actions of $\pi$ and $\vec{r}_\pi$ is the cost
vector of these actions. The matrix $Q_\pi$ is the stripped transition probability matrix of some absorbing Markov
chain. It is well-known that $Q^k_\pi \to 0$ elementwise when $k \to +\infty$ and thus $I-Q_\pi$ is invertible. Since
all elements of $Q_\pi$ are nonnegative, we can multiply both sides of this inequality by $Q_\pi$ and the inequality
still holds row-wise, so for all $k \geqslant 1$ we have $\vec{v} \leqslant Q^k_\pi \vec{v} + Q^{k-1}_\pi \vec{r}_\pi +
\ldots + Q_\pi \vec{r}_\pi + \vec{r}_\pi$, As before, in the limit $k \to +\infty$ we obtain the inequality $\vec{v}
\leqslant (I - Q_\pi)^{-1} \vec{r}_\pi$, from which we conclude that any feasible point is bounded and thus the
objective function is bounded. So, we have verified that the problem \eqref{eq:of} is feasible and bounded, so it has a
solution, which, as we have proved before, is also a solution of the nonlinear equations \eqref{eq:min}.

\section{Proof of Theorem 2}

Let $\vec{v}^*$ be any solution of the system \eqref{eq:min}. For any scheme $\pi$ the solution $\vec{v}^\pi$ satisfies
the system of linear equations $\vec{v}^\pi = Q_\pi \vec{v}^\pi + \vec{r}_\pi$, where $Q_\pi$ and $\vec{r}_\pi$ are
defined above. The optimal solution $\vec{v}^\star$ satisfies the inequality $\vec{v}^\star \leqslant Q_\pi
\vec{v}^\star + \vec{r}_\pi$. Subtracting from this inequality the equality for $\vec{v}^\pi$, we obtain the series of
relations
\begin{equation}
    \vec{v}^\star - \vec{v}^\pi \leqslant Q_\pi (\vec{v}^\star - \vec{v}^\pi) \leqslant \ldots
    \leqslant Q^k_\pi (\vec{v}^\star - \vec{v}^\pi),
\end{equation}
for all $k \geqslant 1$. As we know, $Q_\pi \to 0$ elementwise when $k \to +\infty$. We conclude that $\vec{v}^\star
\leqslant \vec{v}^\pi$ componentwise, which proves the desired relation.

\section{MDP of a quantum repeater}

At any moment in time a state of a repeater can be characterized by a string of numbers, where $0$ always denotes a
segment that is still trying to distribute entanglement, $1$ denotes a segment with entanglement already distributed,
and $i>1$ denotes a group of $i$ successfully swapped segments. The set of all such strings we express by
$\mathcal{S}'_n$. It has been shown in \cite{PhysRevA.100.032322} that the number of such states is given by the odd
Fibonacci number, $|\mathcal{S}'_n| = F_{2n+1}$. As mentioned in the main text, we can combine two states which are
mirror images of each other into one state. This trick reduces the number of states and constraints, thus reducing the
size of the problem and time to solve it. The reduced set of the strings corresponding to the new states is denoted as
$\mathcal{S}_n$. The number of states in $\mathcal{S}_n$ is given by the following theorem.
\begin{thrm}
The size of the set $\mathcal{S}_n$ is given by
\begin{equation}
    |\mathcal{S}_n| = \frac{F_{2n+1} + F_{n+2}}{2}.
\end{equation}
\end{thrm}
\noindent For example, all $|\mathcal{S}_4| = (F_9 + F_6)/2 = 21$ relevant states of a $4$-segment quantum repeater are

\begin{tabular}{p{15mm}p{15mm}p{15mm}p{15mm}p{15mm}}
    0000 & 0001 & 0010 & 0011 & 002 \\
    0101 & 0110 & 0111 & 012 & 020 \\
    021 & 03 & 1001 & 1011 & 102 \\
    1111 & 112 & 121 & 13 & 22 \\
\end{tabular}

\noindent and the terminating state 4, where all four segments have successfully distributed and swapped entanglement.

\begin{proof}
We first prove that the number of symmetric states in $\mathcal{S}'_n$ is the $(n+2)$-th Fibonacci number, $F_{n+2}$.
The set of symmetric states we denote as $S_n \subset \mathcal{S}_n$. It is easy to check that the statement is valid
for $n = 0, 1, 2, 3$. In fact, we have $S_0 = \{\varnothing\}$, $S_1 = \{0, 1\}$, $S_2 = \{00, 11, 2\}$ and $S_3 =
\{000, 010, 101, 111, 3\}$, so $|S_n| = F_{n+2}$ for $n = 0, 1, 2, 3$. We show that the numbers $N_n = |S_n|$ satisfy
the relation
\begin{equation}\label{eq:N}
    N_n = 3N_{n-2} - N_{n-4},
\end{equation}
for $n > 3$. In fact, we can partition the set $S_n$ as
\begin{equation}
    S_n = S'_n \cup S^{\prime\prime}_n,
\end{equation}
where $S'_n$ is the set of symmetric states with the first (and thus the last) element equal to 0 or 1, and
$S^{\prime\prime}_n$ is the set of states which start and end with an element larger than 1. Since $S'_n \cap
S^{\prime\prime}_n = \varnothing$, we have
\begin{equation}
    N_n = |S_n| = |S'_n| + |S^{\prime\prime}_n|.
\end{equation}
Clearly, $|S'_n| = 2N_{n-2}$, since any state from $S'_n$ can be obtained from a state of $S_{n-2}$ in two ways --- by
prefixing and suffixing with 0 or 1. To compute $|S^{\prime\prime}_n|$ note that the states of $S^{\prime\prime}_n$ are
in one-to-one correspondence with the states of $S_{n-2}$ that start and end with 1. It is easier to compute the number
of states of $S_{n-2}$ that end with 0 since this number is just $N_{n-4} = |S_{n-4}|$. We thus have that
$|S^{\prime\prime}_n| = N_{n-2} - N_{n-4}$. Adding the two numbers, we get the relation \eqref{eq:N}. Fibonacci numbers
$F_{n+2}$ also satisfy this relation, as one can easily check from their defining relation $F_{n+1} = F_n + F_{n-1}$.
Since $N_n = F_{n+2}$ for $n = 0, 1, 2, 3$ and for $n > 3$ these numbers satisfy the same recurrence relation, we
conclude that the statement is valid for all $n$.

We can now compute $|\mathcal{S}_n|$. We have
\begin{equation}
    |\mathcal{S}_n| = |S_n| + \frac{|\mathcal{S}'_n| - |S_n|}{2} = \frac{F_{2n+1} + F_{n+2}}{2}.
\end{equation}
\end{proof}

\section{Example of 4-segment repeater}

In our first model (that without including classical communication times on the level of the entanglement swapping) we
count only the time needed to distribute entanglement. For any state with no neighboring ready segments we have no other
choice except waiting when some segments distribute entanglement. If a state has several consecutive ready segments (or
several consecutive groups of segments), then the situation is more interesting since now we have several possibilities.
For example, for $n=4$, consider the state $0111$. We have three possibilities --- either we wait for the first segment
to successfully distribute entanglement, or we try to swap segments 2 and 3, or we try to swap segments 3 and 4. The
first action costs one time unit, while the other two cost nothing. Different choices have different statistical
properties, and in this work we present a method to determine the fastest way to distribute entanglement.

For any state $s \in \mathcal{S}_n$ we introduce the set of actions $\mathcal{A}_s$ that are possible in this state. Any
action $\alpha \in \mathcal{A}_s$ is either waiting for some segments to distribute entanglement (if not all segments
have done it), which we denote by ``distribute'', or trying to swap consecutive groups of ready segments, which we
denote by ``swap $i$ and $i+1$'', where $i$ and $i+1$ are the consecutive ready segments. If there is only one pair of
ready segments, we omit the indices. For example, $\mathcal{A}_{0000} = \{\mathrm{distribute}\}$, since waiting is the
only possibility in this state; $\mathcal{A}_{012} = \{\mathrm{distribute}, \mathrm{swap}\}$; $\mathcal{A}_{1111} =
\{\mathrm{swap}\ 1\ \mathrm{and}\ 2, \mathrm{swap}\ 2\ \mathrm{and}\ 3\}$. Since we work with the compressed state
space, we do not need the action ``swap $3$ and $4$'', because it is identical to the action ``swap $1$ and $2$''.

The union of the actions for all states we denote as $\mathcal{A} = \bigcup_{s \in \mathcal{S}} \mathcal{A}_s$. A scheme
of performing swappings is determined by a choice of an action for any state. Such a choice is referred to as a policy
$\pi$ and is formally defined as a map $\pi: \mathcal{S} \to \mathcal{A}$, such that $\pi(s) \in \mathcal{A}_s$ for all
$s \in \mathcal{S}$. The natural question is how to determine the most efficient policy for a given quantum repeater.
The brute-force approach is to check all of them and choose the best one. For an $n$-segment repeater there is a finite
number of policies $N_n = \prod_{s \in \mathcal{S}} |\mathcal{A}_s|$, so one can try to enumerate and test each of them.
Unfortunately, the number of schemes grows extremely fast. For example, we have $N_2 = 1$, $N_3 = 2$, $N_4 = 384$, $N_5
\approx 4.5 \cdot 10^9$, $N_8 > 10^{315}$, $N_{10} > 10^{2622}$, so this approach is feasible only for $n=3, 4, 5$. We
show that the problem of finding the best scheme can be formulated as a Markov decision problem, reducing it to the
linear programming optimization problem and making it feasible for a wider range of values of $n$.

\section{Classical communication}

The classical communication times on the level of the entanglement swapping
can be also included into this scheme, assuming that the time needed to restart a group of
segments is equal to the number of segments in this group. This significantly increases the size of the problem. The
number of states and constraints of the problem with and without CC is given in the table below for several values of
$n$. \\

\begin{table}[ht]
\begin{tabular}{|r|l|l|l|l|}
\hline
 & \multicolumn{2}{c|}{Without CC} & \multicolumn{2}{c|}{With CC} \\ \hline
$n$ & N. of states & N. of constr. & N. of states & N. of constr. \\ \hline
4 & 20 & 29 & 45 & 56 \\
5 & 50 & 86 & 150 & 206 \\
6 & 126 & 261 & 525 & 791 \\
7 & 321 & 763 & 1,795 & 2,922 \\
8 & 825 & 2,234 & 6,265 & 10,922 \\
9 & 2,134 & 6,424 & 21,877 & 40,502 \\
10 & 5,544 & 18,398 & 76,814 & 150,328 \\ \hline
\end{tabular}
\caption{Size of the MDP with and without CC depending on the number of repeater segments $n$.}
\end{table}

Note that the numbers in the column ``N. of states'' without CC are given by
\begin{equation}
    \frac{F_{2n+1} + F_{n+2}}{2} - 1,
\end{equation}
since we do not include the absorbing state. Explicit expressions for the numbers in the other columns are not known.

\section{The three schemes for $n=4$}

For a uniform description of the optimal actions in each scheme, only those states are presented that allow for more
than a single action. The ``doubling'' scheme, $\pi_0$, is characterized by the following choice of actions:
\begin{alignat*}{3}
    &0011 \to \text{swap} \quad\quad &&1111 &&\to \text{swap 3 and 4} \\
    &0110 \to \text{distribute} \quad\quad &&012 &&\to \text{distribute} \\
    &0111 \to \text{swap 3 and 4} \quad\quad &&112 &&\to \text{swap 1 and 2} \\
    &1011 \to \text{swap 3 and 4} \quad\quad &&021 &&\to \text{swap} \\
\end{alignat*}
The scheme $\pi_1$ is characterized by the following actions:
\begin{alignat*}{3}
    &0011 \to \text{swap 3 and 4} \quad\quad &&1111 &&\to \text{swap 3 and 4} \\
    &0110 \to \text{swap 2 and 3} \quad\quad &&012 &&\to \text{distribute} \\
    &0111 \to \text{swap 3 and 4} \quad\quad &&112 &&\to \text{swap 1 and 2} \\
    &1011 \to \text{swap 3 and 4} \quad\quad &&021 &&\to \text{swap} \\
\end{alignat*}
The scheme $\pi_2$ is characterized by the following actions:
\begin{alignat*}{3}
    &0011 \to \text{swap 3 and 4} \quad\quad &&1111 &&\to \text{swap 3 and 4} \\
    &0110 \to \text{swap 2 and 3} \quad\quad &&012 &&\to \text{swap} \\
    &0111 \to \text{swap 3 and 4} \quad\quad &&112 &&\to \text{swap 1 and 2} \\
    &1011 \to \text{swap 3 and 4} \quad\quad &&021 &&\to \text{swap} \\
\end{alignat*}

Policy $\pi_2$ can be simply interpreted as ``swap as soon as possible'' (as soon as there are ready pairs), whereas for
$\pi_0$ and $\pi_1$ this is not always the case (e.g. see the state 012). However, note that in general, for larger $n$,
there will be many different schemes, each optimal in a corresponding small region of the unit square $(p, a)$ and
having a tiny advantage over each other in neighboring regions; so, in the general case, a simple interpretation as
``swap as soon as possible'' for the optimal schemes is not applicable. Further note that there is always strictly only
one action in any given state and the optimization finds the best action in each state. If in some state two or more
actions have the same optimal behavior one may choose any of them randomly.

For $n=4$, policy $\pi_1$ swaps as soon as possible on the lowest level (on the level of connecting the initial
segments), unlike $\pi_0$, see the state $0110$. A possible explanation of these different optimal policies is the
following. The higher the chance for a failed swapping attempt (smaller $a$), the better it is to directly swap only
small groups of segments, or even only groups of segments that belong to ``independent'' smaller sub-repeaters on the
lower levels. Trying to connect these sub-repeaters as soon as possible is the best strategy for larger $a$.

As a final remark we notice that for a given scheme some states may not be reachable from the initial state, e.g. see
$021$ for $\pi_0$ (``doubling''), which nonetheless is listed above. As the optimal scheme is not known beforehand,
inequalities have to be included for all states. In practice, once the optimal solution has been found, we would care
only about those states reachable from some initial state (which is $0000$ for $n=4$ in our model) and we can ignore
some parts of the information provided by the optimal solution. If the system would start in some different initial
state, the optimal policy would also become available. Would we know beforehand what states are unreachable, the
optimization problem would become smaller, but this information is only accessible afterwards together with the
solution. Nonetheless, reachability of states is a meaningful element of the finally obtained optimal policies.

\section{Supplemetary figures}

\begin{figure*}
    \includegraphics{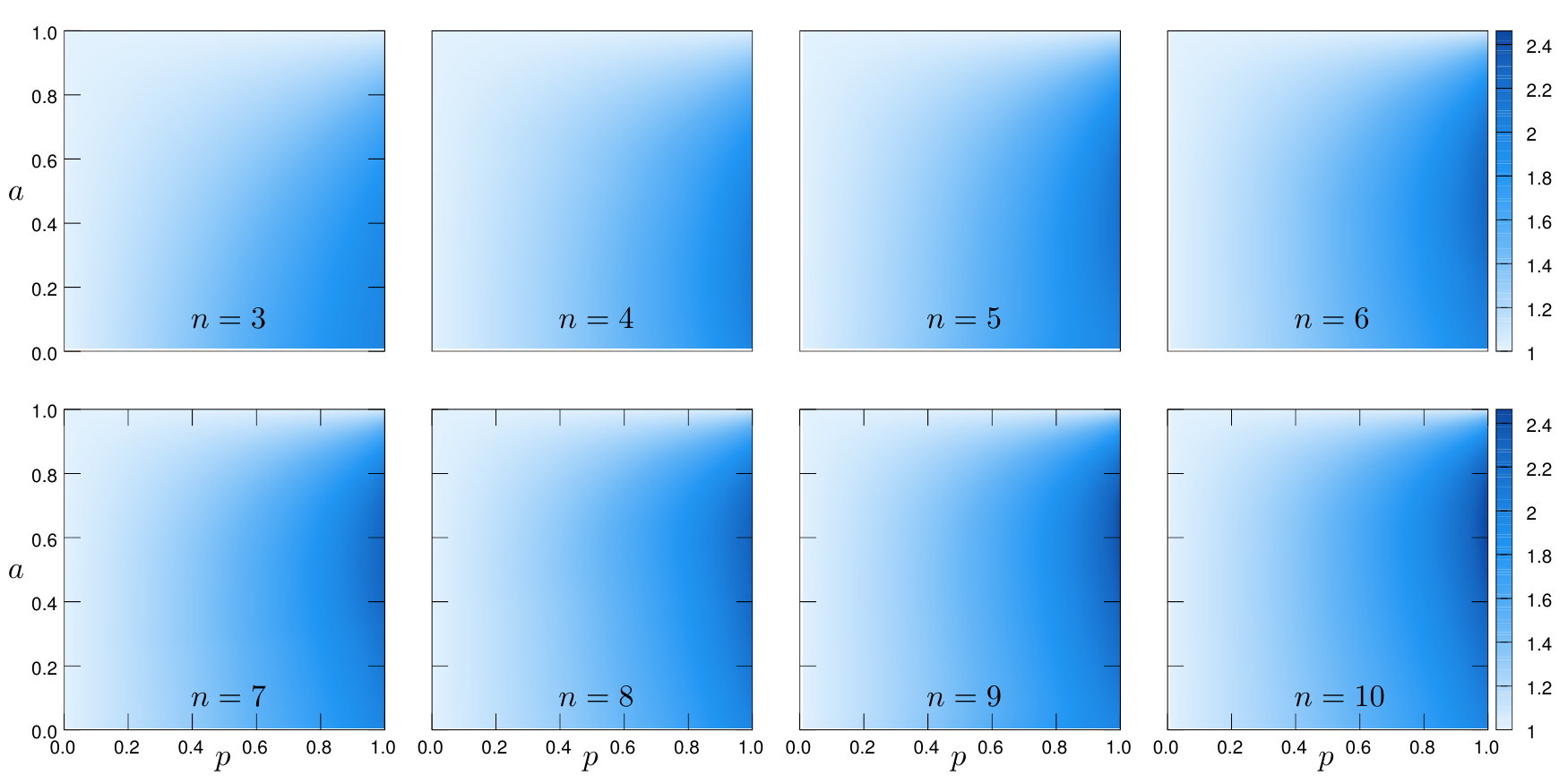}
    \caption{Ratio between the optimal raw waiting times with CC and without for different numbers of segments $n$. The
    relative impact of CC is small for small $p$ and becomes maximal for $p$ approaching one and $a$ at around one
    half.}
    \label{fig:CC4}
\end{figure*}

The two additional figures below show a comparison between optimal and ``doubling'' waiting times for the simplest model
without all CC times and between the optimal waiting times with and without CC times for various repeater sizes. From
the latter comparison, we can infer that the impact of CC grows with an increasing number of repeater segments. In
Fig.~\ref{fig:CC4}, this is not always easy to see, but, for example, for $n=9$ the maximal impact (ratio) is 2.41, for
$n=10$ it is 2.46.

\begin{figure}
    \includegraphics{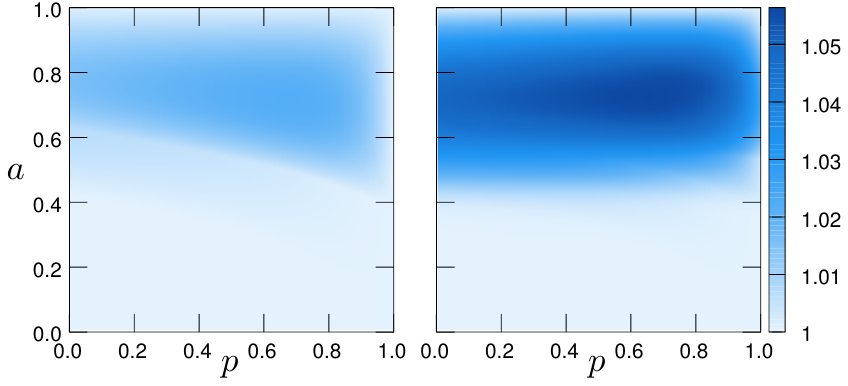}
\caption{Ratio of the ``doubling'' to the optimal waiting time for $n = 4$ (left) and $n = 8$ (right), model without CC.}
\end{figure}

\end{document}